\documentclass[a4paper]{article}

\usepackage[colorlinks=true, citecolor=green, linkcolor=black, urlcolor=blue]{hyperref}
\usepackage[font=footnotesize]{caption}
\usepackage{authblk}
\usepackage{graphicx, wrapfig, rotating}
\usepackage{dirtytalk}
\usepackage{lineno}
\usepackage{enumitem}

\graphicspath{{.//code_letter_nature/code_letter_nature_files/figure-latex/}}

\usepackage[style=nature, url=false, doi=true, isbn=false, sorting=none, maxcitenames=2, maxbibnames=30, giveninits=true]{biblatex}

\renewbibmacro{in:}{}

\DeclareFieldFormat{pages}{#1}

\begin{document}

\pagenumbering{arabic}
\date{}
\title{Are the results of the groundwater model robust?}
\author[1,2]{Arnald Puy\thanks{Corresponding author}}
\author[3]{Emanuele Borgonovo}
\author[4]{Samuele Lo Piano}
\author[2]{Andrea Saltelli}

\affil[1]{\footnotesize{\textit{Department of Ecology and Evolutionary Biology, M31 Guyot Hall, Princeton University, New Jersey 08544, USA. E-Mail: apuy@princeton.edu}}}

\affil[2]{\footnotesize{\textit{Centre for the Study of the Sciences and the Humanities (SVT), University of Bergen, Parkveien 9, PB 7805, 5020 Bergen, Norway.}}}

\affil[3]{\footnotesize{\textit{Department of Decision Sciences and BIDSA, Bocconi University,
Via Roentgen 1, 20136, Milano, Italy}}}

\affil[4]{\footnotesize{\textit{University of Reading, School of the Built Environment, JJ Thompson Building, Whiteknights Campus, Reading, RG6 6AF, United Kingdom}}}

\maketitle


\textbf{\textcite{DeGraaf2019} suggest that groundwater pumping will bring 42--79\% of worldwide watersheds close to environmental exhaustion by 2050. We are skeptical of these figures due to several non-unique assumptions behind the calculation of irrigation water demands and the perfunctory exploration of the model's uncertainty space. Their sensitivity analysis reveals a widespread lack of elementary concepts of design of experiments among modellers, and can not be taken as a proof that their conclusions are robust.}

\textcite{DeGraaf2019} estimate when and where groundwater pumping will cause a critical decrease in streamflow globally. Since irrigation agriculture is responsible for $\sim$70\% of all groundwater pumped worldwide, the assumptions behind the modelling of irrigation water demands strongly ground \textcite{DeGraaf2019} 's results. Should they had systematically accounted for the uncertainties embedded in
their modelling exercise, we believe that their conclusions would have been substantially different.

Our first concern derives from their use of the spatially-distributed, global hydrological model PCR-GLOBWB \cite{Wada2014}. This model follows FAO guidelines to provide estimates of irrigation water demands as a function of the irrigated area and the crop water requirements. The latter is partially formalized with the classic crop evapotranspiration ($ET_c$) equation 

\begin{equation}
ET_c=k_cET_0
\label{eq:evapotranspiration}
\end{equation}

where $k_c$ is the crop coefficient and $ET_0$ is the reference evapotranspiration. We would like to raise attention on the following points: 
\begin{itemize}
\item PCR-GLOBWB uses $k_c$ values reported in \textcite{Allen1998}, which were obtained for specific locations under specific management practices. It is known that extrapolation of $k_c$ values to different settings might lead to biased water demand computations \cite{Jagtap1989}, and that a 10\% uncertainty in $k_c$ values can modify irrigation water requirements up to 15\% \cite{Satti2004}.

\item PCR-GLOBWB computes $ET_0$ with the Penman-Monteith equation, one of the $\sim$50 formulae available \cite{Lu2005}.  $ET_0$ values are highly sensitive to the selection of the formula, which can vary the results by up to 600 mm \cite{Kingston2009}. The Priestley Taylor or the Kimberly Penman methods, for instance, can yield much higher $ET_0$ values than the Penman-Monteith \cite{Weis2008}.

\item Evaporation and crop transpiration are simulated by PCR-GLOBWB over the irrigated areas documented in the Global Map of Irrigation Areas (GMIA) \cite{Siebert2005}. We are aware of at least five more products informing on irrigated areas apart from the GMIA \cite{Meier2018}, with differences on the extension of irrigation that can be as large as one order of magnitude at the country level.
\end{itemize}

A detailed account of the uncertainties hidden in the computation of irrigation water demands in PCR-GLOBWB is beyond the scope of this letter. To our knowledge, no study has thoroughly conducted a global sensitivity analysis of PCR-GLOBWB (see \textcite{Wada2014} for an approach to climatic uncertainties and a scenario analysis of surface water and groundwater withdrawal). The same applies to the groundwater flow model that \textcite{DeGraaf2019} couple with PCR-GLOBWB, which is based on MODFLOW \cite{DeGraaf2017}. \textcite{DeGraaf2019} are thus building their study on a combination of models whose sensitivity to alternative yet perfectly valid assumptions, as well as to parametric uncertainty, remains significantly unexplored.

\begin{figure}[!ht]
\centering
\includegraphics[keepaspectratio, width=\textwidth]{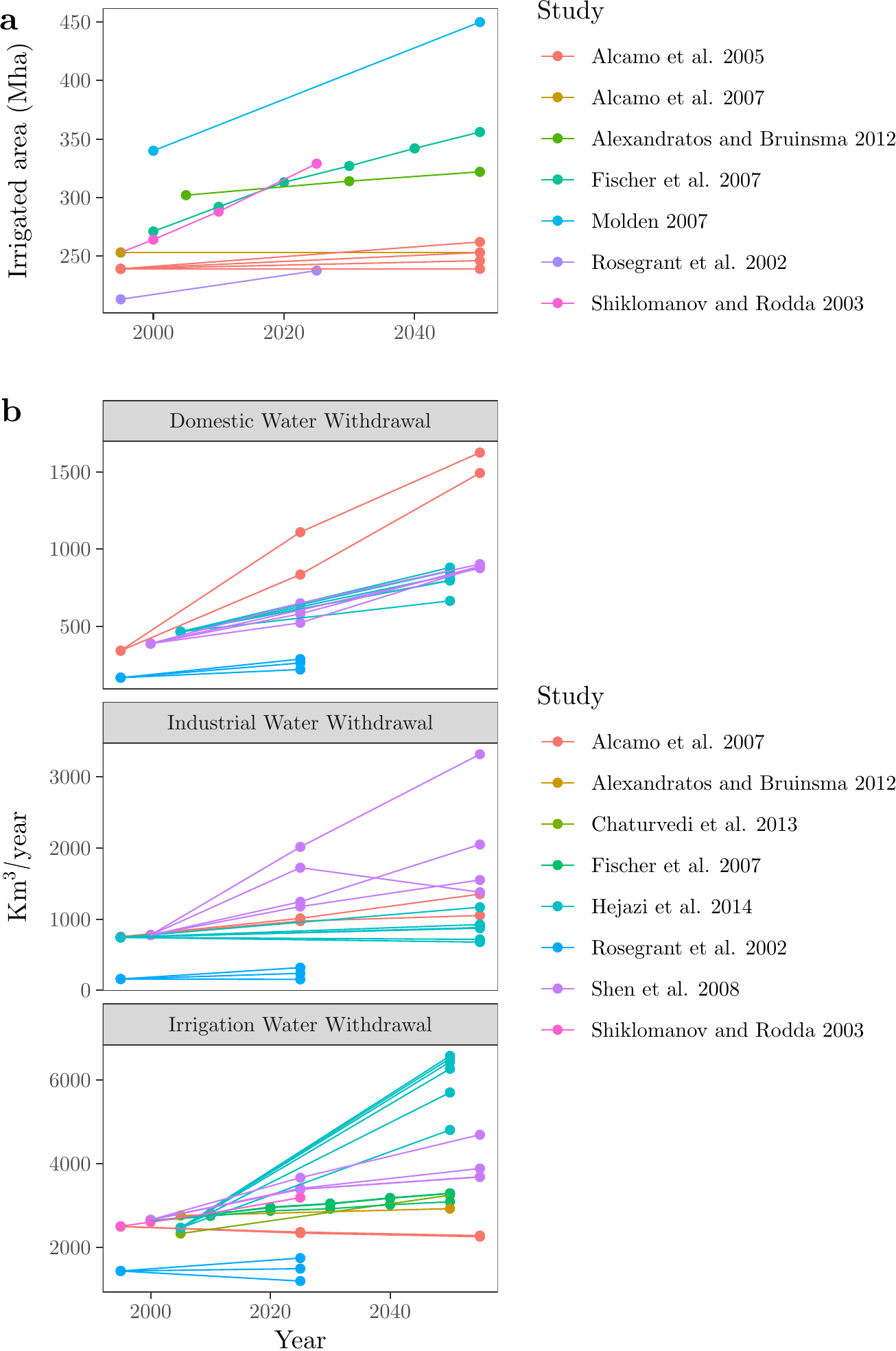}
\caption{Some available global projections of a) irrigated areas, and b) water demand across sectors. Lines with the same colour in the same plot reflect the behaviour of the output of interest in different scenarios contemplated by the authors. Full references for the studies are provided in the Supplementary Information File.}
\label{fig:plot}
\end{figure}

Our second concern derives from \textcite{DeGraaf2019}'s  assumption of irrigated areas and industrial/domestic water demands remaining constant from 2010 to 2100. This exclusive reliance on a \say{business-as-usual} scenario appears as a methodological choice rather opposed to many available studies suggesting a significant increase of irrigated areas and water demand in all sectors between 2000--2050 (Figure~\ref{fig:plot}, see also Supplementary Materials). The extra pressure on water resources put by larger irrigated areas is also overlooked by \textcite{DeGraaf2019}, which only consider climate change as a factor affecting future irrigation water withdrawals.  

Our third concern is on the adequacy of \textcite{DeGraaf2019}'s sensitivity analysis, which plays a prominent role in their paper [the word \textit{sensitiv}* is used 24 times (3 x \textit{sensitive}, 19 x \textit{sensitivity}, 1 x \textit{sensitivities})]. The authors explore how sensitive is their model to uncertainties in three different sub-structures, each one isolated from the rest:

\begin{enumerate}
\item Values in hydraulic conductivity, drainage level and river bed conductance.
\item The definition of environmental flow limits.
\item The climate model selected.
\end{enumerate} 

For each structure, they vary the parameter of interest, an approach \textcite{DeGraaf2019} describe as a one-at-a-time (OAT) sensitivity analysis.

This approach is unable to provide any useful information: firstly, because sensitivities can not be fully explored by compartmentalizing model structures. By overlooking how their model output is affected by simultaneous variations in (1, 2, 3), \textcite{DeGraaf2019} miss interaction effects among the parameters of the different compartments. Interactions are paramount in non-additive models, which is likely the case with \textcite{DeGraaf2019}'s due to the presence of exponential terms in the sub-structures. In that context, a non-compartmentalized OAT across the five uncertain parameters/structures in (1, 2, 3) would have allowed the authors to explore only 16\% of the model's uncertainty space \cite{Saltelli2010b}. Given the much higher dimensionality of hydrological models and the piecewise sensitivity analysis of \textcite{DeGraaf2019}, there are reasons to suspect that the authors only explore a minimal or negligible part of the input space (in ten dimensions, for instance, that would be 0.002\%).

Even with their piecewise approach to sensitivity, \textcite{DeGraaf2019} should have been able to compute main effects plus interactions in 1), as they rely on a $3^2$ full factorial design. However, this information is not provided.

Secondly, in 3), \textcite{DeGraaf2019} actually conduct a scenario analysis: a point in the input parameter space is combined with three different global climate models. This approach is also questionable: are the scenarios meaningful, i.e. shared references for the entire modeling community?. Even if they are, \textcite{DeGraaf2019} are only considering the isolated climatic effect, not the effects of combining climate models with other points in the input parameter space. Such an approach of scenario decomposition in the climate sciences is much more defensible and is exemplified by \textcite{Marangoni2017}.

Thirdly and lastly, \textcite{DeGraaf2019} might be right when they regard a Monte-Carlo sensitivity analysis unaffordable due to its high computational demand. However, there are well-established alternatives that would have allowed them to better explore the full uncertainty space: variance-based measures of sensitivity can be applied to well-constructed and efficient emulators [i.e. based on Gaussian processes or on polynomial chaos expansions (PCE)]. They allow for error quantification and the computation of sensitivity indices. Even OAT would have allowed them to compute interactions by only doubling the number of model runs \cite{Borgonovo2011}. These tools could be complemented by a series of triggers that enact the different modelling approaches discussed above. In this way, one could assess the sensitivity of the model output to uncertainties in the model structures available. 

Given \textcite{DeGraaf2019}'s approach to uncertainties, it is hard to ascertain whether their results are too optimistic or too pessimistic. We argue that they are simply unreliable: the timing of groundwater exhaustion and the number of watersheds reaching flow limits by 2100 might simply be the product of several concatenated and non-unique modelling assumptions. Due to the important role that hydrological models play in guiding our approach to water security and environmental welfare, their cascade of uncertainties should be accounted for and be systematically investigated. The risks of not doing so, in contrast, are too certain to be ignored \cite{Pilkey2009}.

\printbibliography

\section*{Competing interests}
A.P. has worked on this paper on a Marie Sk\l{}odowska-Curie Global Fellowship (grant number 792178).

\section*{Author contributions}
A.P. wrote the paper, with contributions from E.B., S.L.P. and A.S. 

\end{document}


\pagenumbering{arabic}
\date{}
\title{Are the results of the groundwater model robust? \\ \vspace{2mm} \large{Supplementary Materials}}
\author[1,2]{Arnald Puy\thanks{Corresponding author}}
\author[3]{Emanuele Borgonovo}
\author[4]{Samuele Lo Piano}
\author[2]{Andrea Saltelli}

\affil[1]{\footnotesize{\textit{Department of Ecology and Evolutionary Biology, M31 Guyot Hall, Princeton University, New Jersey 08544, USA. E-Mail: apuy@princeton.edu}}}

\affil[2]{\footnotesize{\textit{Centre for the Study of the Sciences and the Humanities (SVT), University of Bergen, Parkveien 9, PB 7805, 5020 Bergen, Norway.}}}

\affil[3]{\footnotesize{\textit{Department of Decision Sciences and BIDSA, Bocconi University,
Via Roentgen 1, 20136, Milano, Italy}}}

\affil[4]{\footnotesize{\textit{University of Reading, School of the Built Environment, JJ Thompson Building, Whiteknights Campus, Reading, RG6 6AF, United Kingdom}}}

\maketitle


Here we provide full references for the studies mentioned in Fig. 1 \cite{Alcamo2005, Alcamo2007, Alexandratos2012, Fischer2007, Molden2007, Rosegrant2002, Shiklomanov2003, Chaturvedi2013, Hejazi2014, Shen2008}.

\printbibliography